\documentclass[rnote]{aa} 
\usepackage{graphicx}
\usepackage{natbib}
\bibstyle{aa}

\usepackage{txfonts}
\begin{document}
   \title{Redshift Limits of BL Lacertae Objects from Optical Spectroscopy}

   \author{J. D. Finke\inst{1,2}\fnmsep\thanks{NRC/NRL Research Associate}
          \and
          J. C. Shields\inst{1}
	  \and
	  M. B\"ottcher\inst{1}
	  \and
	  S. Basu\inst{1}
          }

   \offprints{J. D. Finke, \email{jfinke@ssd5.nrl.navy.mil}}

   \institute{
     Astrophysical Institute, Department of Physics and Astronomy, \\
     Ohio University, Athens, Ohio, 45701, USA  \\
         \and
     Current Address:  \\
     Code 7653, US Naval Research Laboratory, \\ 
     Washington, DC, 20375, USA \\
     \email{jfinke@ssd5.nrl.navy.mil}
             }

   \date{}

 
  \abstract
{BL Lacertae objects have been the targets for numerous recent multiwavelength
campaigns, continuum spectral variability studies, and theoretical
spectral and variability modeling.  A meaningful interpretation
of the results of such studies requires a reliable knowledge of the
objects' redshifts; however, the redshifts for many are still unknown
or uncertain.  }
{Therefore, we hope to determine or constrain the redshifts of 
six BL Lac objects with unknown or poorly known redshifts.  }
{Observations were made of these objects with the MDM 2.4 m
Hiltner telescope.  Although no spectral features were detected, and
thus no redshifts could be measured, lower redshift limits were assigned
to the objects based on the expected equivalent widths of absorption
features in their host galaxies.  Redshifts were also estimated for
some objects by assuming the host galaxies are standard candles 
and using host galaxy apparent magnitudes taken from the literature.
}
{The commonly used redshift of $z=0.102$ 
for \object{1219+285} is almost certainly wrong, while
the redshifts of the other objects studied remain undetermined.  }
{}


   \titlerunning{Redshift Limits of BL Lacertae Objects}
   \authorrunning{Finke et al.}
   \maketitle
%

\section{Introduction}

Blazars are among the most violent, highly variable astrophysical
high-energy phenomena, with rapidly-varying emission from radio
through $\gamma$-ray energies.  They are thought to consist of relativistic
jets from supermassive black holes,
closely aligned with our line of sight.  

BL Lacertae (BL Lac) objects, a subclass of blazars, are defined by their
quasar-like continuum and the weakness or absence of broad
emission lines in their optical spectra.  These spectra are thought
to be dominated by nonthermal emission from the highly relativistic
jet, masking
the contributions of the stars in the host galaxy, and
line emission from gas clouds near the supermassive black hole.

There are various models for the source of the nonthermal emission
from blazars.  Of critical importance in
distinguishing between the models is setting the energy scale,
on which many parameters (e.g., the jet's speed, density,
magnetic field, etc.) depend.  This can only be done if the objects'
redshifts, and hence distances, are known.

Many well-observed BL Lac objects have unknown or poorly known
redshifts based on $\sim$1--3 emission or absorption lines in
low signal-to-noise ($S/N$) spectra.  In
several cases, a poorly determined redshift of a BL Lac is repeatedly
cited throughout the literature until its reliability is no longer
questioned.  Thus, a project was undertaken to obtain the optical 
spectra of several of these objects at the MDM Observatory with the 
goal of obtaining redshift estimates or improved constraints.

Unfortunately, no definitive spectral features
were revealed, and thus no definitive redshifts could be assigned.
However, using the method of \citet{sbarufatti06}, we were able to
estimate a minimum redshift based on the expected equivalent widths
of absorption features in the host galaxy.  Also, assuming
the host galaxies are standard candles 
\citep[see, e.g.,][]{nilsson03}, we estimated the redshifts of
several objects based on the observed magnitude of the host galaxies,
found from the literature, based on the method of 
\citet{sbarufatti05}.


\section{Observations and Data Analysis}
\label{data}

\subsection{Observations}
\label{observations}

\begin{table*}
\caption{Blazar observations with the Hiltner telescope.\label{obstable}}
\centering
\begin{tabular}{rrrccr}
\hline\hline
Object & RA (J2000) & Dec (J2000) & Setting & Exp. Time [sec] & Obs. Date (UT)\\
\hline
\object{0219+428} & 02:22:39.6 & +43:02:08 & blue & 4800 & 29 Nov. 2005 \\
                  &            &           & green & 7200 & 29 Nov. 2005 \\
                  &            &           & blue & 3600 & 30 Nov. 2005 \\
                  &            &           & red & 10800 & 30 Nov. 2005 \\
\object{0716+714} & 07:21:53.4 & +71:20:36 & green & 7200 & 29 Nov. 2005 \\
         &            &             & blue & 5400 & 30 Nov. 2005 \\
\object{1011+496} & 10:15:04.1 & +49:26:01 & blue & 7200 & 26 Mar. 2006 \\
\object{1055+567} & 10:58:37.7 & +56:28:11 & blue & 7200 & 26 Mar. 2006 \\
\object{1219+285} & 12:21:31.7 & +28:13:59 & blue & 5400 & 26 Mar. 2006 \\
\object{1426+428} & 14:28:32 & +42:40:21 & blue & 5090 & 26 Mar. 2006 \\
\hline
\end{tabular}

\end{table*}

Spectroscopic observations of 
six BL Lac objects were taken with the MDM 2.4 m Hiltner telescope
in November 2005 and March 2006.  
Arc lamps and one standard star per night per wavelength setting were
also observed for wavelength and spectrophotometric calibration, 
respectively.  The CCDS spectrograph was used with a slit 
width of $1.5\arcsec$ and the 350 grooves/mm grating.  
The grating provides a wavelength range of 1592 \AA.
Three different observing settings were
used:  $\sim 4000-5500$ \AA\ (hereafter referred to as the blue setting), 
$\sim 5500-7000$ (green setting) and $\sim 7000-8500$ \AA\ (red setting).
For the red and green settings, the LG-400 order-blocking filter was
used.  Unfortunately, weather conditions did not allow observation 
of all the sources with all of the settings.  Seeing during
the observations was in the range $\sim 1-3$\arcsec\ with an average of 
$\sim 1.5$\arcsec.  On 29 November conditions
were nearly photometric; on the other nights, thin cirrus drifted in and 
out of the field. On 29 November the seeing was $\sim 2\arcsec$, and 
the slit was not consistently aligned with the parallactic angle, 
leading to noticeable flux losses from atmospheric
differential refraction.  A summary of 
the observations can be found in Table \ref{obstable}.

The spectra were reduced with IRAF\footnote{IRAF (Image Reduction
and Analysis Facility) is distributed by the National Optical
Astronomy Observatory which is operated by the Association
of Universities for Research in Astronomy, Inc., under
cooperative agreement with the National Science Foundation.}
using standard methods.  For each object and
grating setting, 3-4 spectra were taken, which were then 
averaged into a single spectrum.

\subsection{Redshift Limit Procedure}
\label{procedure}

The procedure for finding the minimum redshift of the BL Lac objects
is described by \citet{sbarufatti06} We refer the reader to 
that work, and point out that their eq. (1) contains a typographical 
error \citep{finke07} and should be
\begin{equation}
\label{EWeqn}
EW_{obs} = \frac{ (1 + z) EW_0 } {1 + \rho (\lambda) / A(z)}
\end{equation}
where $EW_0$ is the equivalent width of a line in the rest frame of 
the galaxy, $\rho$ is the AGN flux to host galaxy flux ratio and
$A(z)$ is the aperture correction, which takes into account the
fact that not all of the galaxy's emission will be in the 
extraction aperture.

To determine the lower limit, one needs two observations:  
the minimum possible $EW$ one could observe with the spectrum, 
and the appearent magnitude of the object (both these things 
are listed in Table \ref{resulttable}).
For the objects \object{1011+496} and \object{1055+567}, 
photometric observations on the
same night were taken with the McGraw-Hill 1.3 m telescope.  
For the rest of the
spectra, $m_B$ was determined by integrating the spectrum weighted by 
the response function for the $B$ filter using the IRAF utility
SBANDS.  The magnitudes of 
\object{1011+496} and \object{1055+567} taken
from photometric observations were compared with the magnitudes derived
from the spectra and found to vary by less than 0.5 mag.  
We compared the standard star observation on 26 March  
with an exposure of that star aborted due to losing the guide star by 
heavy cloud cover.  The difference was $\sim 0.52$ magnitudes.  
We consequently adopted 0.5 mag as a reasonable estimate of the 
uncertainty in spectroscopic magnitudes for the other targets.  

\section{Results}
\label{results}  

\begin{figure*}
\centering
\includegraphics[width=13cm]{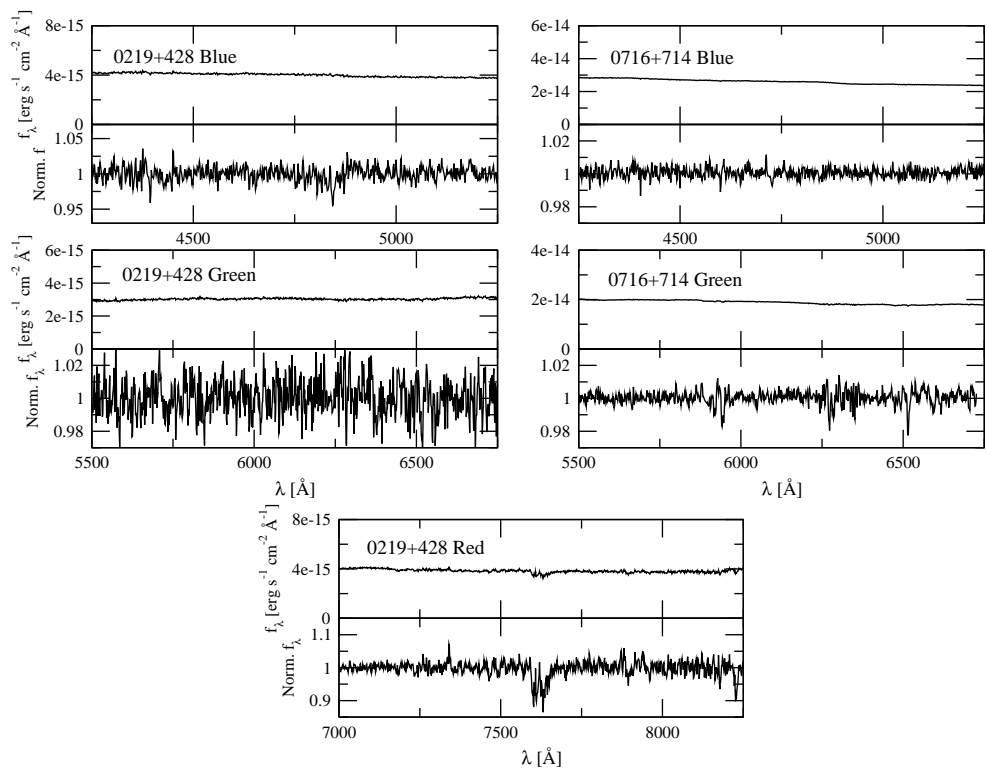}
\caption{Blazar spectra taken in 2005.}
\label{spec1}
\end{figure*}

\begin{figure*}
\centering
\vspace{1.5cm}
\includegraphics[width=13cm]{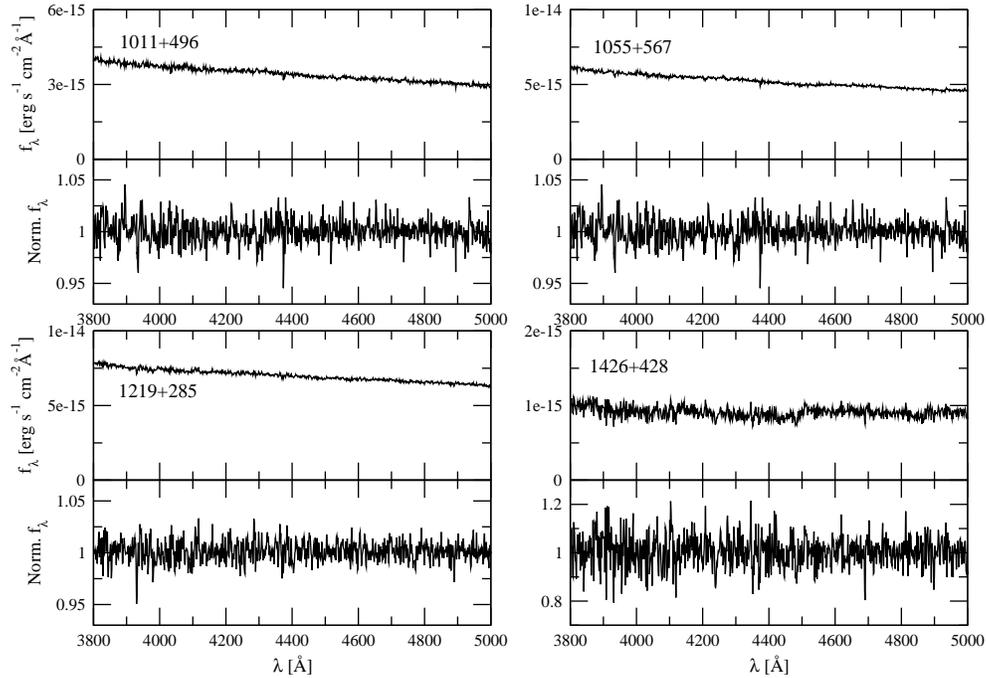}
\caption{Blazar spectra taken in 2006.}
\label{spec2}
\end{figure*}

Flux-calibrated and normalized spectra for our sources are seen in 
Figs \ref{spec1} and \ref{spec2} 
and results are sumarized in Table \ref{resulttable}.  
In a few instances the spectra show
residual structure that we attribute to imperfect flux 
calibration (e.g., at $\sim4880$ \AA\ in the 2005 spectra 
and $\sim4500$ \AA\ in the 2006 spectra), and variations 
of slope between grating settings of similar origin.  
Low-frequency variations of this type do not significantly 
impact our analysis.

\begin{table*}
\begin{center}
\caption{Results of blazar observations.\label{resulttable}}
\begin{tabular}{rccccc}
\hline\hline
Object & $m_B$ & $S/N$ & $EW_{min}$ [\AA] & $z_{spec}$ & $z_{phot}$  \\
\hline
\object{0219+428} & $15.35 \pm 0.50$ & 90 & 0.87 & $\geq$0.096 & 0.321\\ 
\object{0716+714} & $13.39 \pm 0.50$ & 270 & 0.27 & $\geq$0.070 & -\\
\object{1011+496} & $15.80 \pm 0.02$ & 50 & 1.001 & $\geq$0.134 & $0.213 \pm 0.041$\\
\object{1055+567} & $15.60 \pm 0.02$ & 70 & 0.79 & $\geq$0.136 & -\\
\object{1219+285} & $14.88 \pm 0.50$ & 90 & 0.43 & $\geq$0.104 & $0.161 \pm 0.035$\\
\object{1426+428} & $17.10 \pm 0.50$ & 10 & 3.17 & $\geq$0.106 & $0.132 \pm 0.030$\\
\hline
\end{tabular}

\end{center}
\end{table*}

A brief description of the previous observations and results for  
individual objects are discussed below.  All error bars quoted are
$1\sigma$ error bars.

{\bf \object{0219+428} (3C 66A)} has been the target of a recent 
multiwavelength campaign \citep{boett05} and has been extensively
observed in the radio to $\gamma$-rays.  Its optical spectrum 
was first observed by \citet{wills74}, who found it to be flat 
and featureless.  Observations by \citet{miller78} with 
the Lick 3 m Shane reflector revealed an emission feature 
at 4044 \AA, which they identified as \ion{Mg}{ii} 2800, giving the 
object a redshift of $z=0.444$.  However, the feature is located
in a region where it is confused with
 telluric H$_2$O absorption, and 
the authors did not consider it reliable.  {\it International 
Ultraviolet Explorer} observations detected a feature at 1750 \AA\  
which could be Ly$\alpha$ emission at a redshift of $z=0.444$, but
the redshift of 
\object{0219+428} is still far from certain \citep{lanzetta93}.  
Its host galaxy was marginally resolved by \citet{wurtz96}, and
found to have a magnitude of $r_{Gunn}=19.0$.  Converting to 
Johnson $R$ using the prescription of \citet{kent85} yields 
$m_R = 18.43$, and thus a photometric redshift of $z = 0.321$.  
\citet{wurtz96} do not provide an error estimate for the magnitude;
however, since it was only marginally resolved, the error can
be assumed to be high.  Our spectrum's wavelength range does not cover the 
range of the \citet{miller78} feature, so we are unable to confirm or 
refute their
detection; however, we were able to constrain \object{0219+428}'s redshift 
to $z\geq0.096$.  The \object{0219+428} green spectrum was likely 
affected by slit losses due to atmospheric dispersion, which 
would explain the difference in slope from the red and blue 
spectra.  Although we removed most telluric features, 
we were unable to fully remove the A band feature (7594 \AA) 
in the red spectrum.

{\bf PKS \object{0716+714}} has been detected in X-rays and $\gamma$-rays
\citep[see, e.g., ][]{foschini06}.  High $S/N$ spectra 
taken with the KPNO 2.1 m and MMT 6.5 m telescopes 
are flat and featureless \citep{rector01}.  
Our observations also did not reveal 
any spectral features, and we constrained its redshift to $z\geq0.070$; 
despite its high $S/N$, its brightness does not allow for a 
higher constraint.  
Its host galaxy is unresolved, and \citet{sbarufatti05} used this
to constrain its redshift to $z \geq 0.52$.

{\bf \object{1011+496}} has been detected by EGRET \citep{thompson95},
BeppoSAX \citep{donnato05} and MAGIC
\citep{albert07}.  Its last published spectrum was obtained with 
the McDonald Observatory 2.7 m telescope 
\citep{machalski91}.  Their spectrum was from 3200 \AA\ to 
6000 \AA\ and included an unidentified feature at $\sim 3700$ \AA.  
Another spectrum with the KPNO 2.1 m telescope from 4400-6500 \AA\
showed no emission or absorption lines.  Neither spectrum had a 
$S/N>5$. Its redshift is usually 
quoted as $z = 0.20$, as this is the redshift of the nearby cluster, 
A950, to which \object{1011+498} is presumed to belong \citep{leir77}.  
With our higher $S/N$ spectrum we were also unable to detect 
any spectral features; however, we could constrain its redshift 
to $z \geq 0.134$.  With a resolved host galaxy magnitude of 
$m_R = 17.30$ from HST observations \citep{urry00}, 
its photometric redshift can 
be estimated to be $z = 0.213 \pm 0.041$, 
in agreement with it being part of the cluster A950.  
\object{1011+496} has never had an 
optical spectrum published beyond 6500 \AA.  

{\bf \object{1055+567}} has also been detected by EGRET \citep{thompson95}
and BeppoSAX \citep{donnato05}.  \citet{marcha96} report a 
redshift of $z=0.410$ based on probable detection of the
[\ion{O}{iii}] doublet in MMT spectra.  A measurement by \citet{bade98} 
using the WHT, obtained with the source at a comparable flux 
level, does not appear to confirm the [\ion{O}{iii}] emission and the 
authors instead estimate $z=0.144$ based on \ion{Na}{i} D absorption 
and blended H$\alpha$ + [\ion{N}{ii}] emission.  However, both features 
are very weak, and the putative emission feature is additionally 
suspect since it sits on the wing of the atmospheric A-band 
absorption feature.  Our spectrum does not cover the 
wavelength range of these features, and we can only constrain 
the redshift to be $z\geq0.136$.  This object
has not had an observation 
of its host galaxy published, thus we cannot estimate a photometric 
redshift.  It should be noted that, although the $z=0.144$ value 
is cited more often in the literature (and is quoted as such in 
the Simbad database), the $z=0.410$ redshift is still used as well.

{\bf \object{1219+285} (W Comae)} has been detected by EGRET \citep{sreekumar96} 
and is considered a promising target for Very High Energy (VHE)
 $\gamma$-rays by instruments such as VERITAS or MAGIC.  
\citet{weistrop85} 
performed spectroscopy on the object with the 4 m KPNO telescope and 
estimated a redshift of $z=0.102$ based on [\ion{O}{iii}] and H$\alpha$.  
However, the spectrum shows strong residuals due to sky subtraction 
and possibly other problems, and the authors acknowledge the line 
identifications are uncertain.  Our high $S/N$ spectrum did not
reveal any spectral features, but we did not observe in a range that 
would allow us to confirm the features detected by \citet{weistrop85}.  
\citet{nesci01} published a spectrum of the object with the 2.6 m
Byurakan Observatory telescope but were unable
to confirm the observation of [\ion{O}{iii}] detected by \citet{weistrop85}.
The host galaxy of \object{1219+285} was resolved by \citet{nilsson03}, and
they found its magnitude to be $m_R = 16.60 \pm 0.10$.
Based on this measurement, we
estimate its photometric redshift to be $z = 0.161 \pm 0.035$, 
a considerable discrepancy with the spectroscopic value of
\citet{weistrop85}.
We could spectroscopically constrain the redshift of \object{1219+285} 
to $z\geq0.104$.  It therefore seems unlikely that the \citet{weistrop85}
redshift of $z=0.102$ is correct.

{\bf \object{1426+428}} has been detected at VHE $\gamma$-rays by CAT and
HEGRA \citep{djannati02,petry00} and has been the target of 
multiwavelength campaigns \citep{horns03}.  Its only reported optical 
spectrum was published by \citet{remillard89}.  Their highest $S/N$ 
spectrum ($S/N \sim 10$) with the MDM 1.3 m telescope yielded 
$z=0.129$ from marginal detections of \ion{Mg}{i} and \ion{Na}{i} 
at $\sim 5800$ and $\sim 6650$ \AA.  
Unfortunately, we were not able to achieve a 
higher $S/N$ spectrum, nor were we able to observe at a wavelength 
above 5500 \AA, thus we could not confirm this observation; we could 
only constrain its redshift to $z\geq0.106$.  
The host galaxy of \object{1426+428} was 
resolved by \citet{urry00} and found to have $m_R=16.14$, leading 
to a photometric redshift of $z=0.132\pm0.030$.  Our measurements 
are consistent with the previous redshift claims.  

\section{Summary}
\label{summary}

The spectra of six BL Lac objects with poorly known or 
unknown redshifts have been obtained.  
For several objects, these spectra have higher 
$S/N$ than any previously published.  Based on this papers' results, 
the commonly used redshift of $z=0.102$ 
for \object{1219+285} is almost certainly wrong.
The redshifts of the other objects studied remain undetermined.

\begin{acknowledgements}
We thank the anonymous referee for helpful comments.
\end{acknowledgements}

\end{document}